\documentclass[%
 reprint,
 superscriptaddress,
nofootinbib,
 amsmath,amssymb,
 aps,
]{revtex4-2}

\usepackage{graphicx}
\usepackage{dcolumn}
\usepackage{bm}
\usepackage{hyperref}
\usepackage{xcolor}

\usepackage[justification=raggedright]{caption}
\usepackage{subcaption}
\begin{document}

\title{Combining non-parametric quantum states and MERA tensor networks for ground-state optimization}

\newcommand{\affiliationibm}{IBM Quantum, IBM Research Europe - Zurich, S\"aumerstrasse 4, 8803 R\"uschlikon, Switzerland}
\newcommand{\affiliationepfl}{Institute of Physics, \'Ecole Polytechnique F\'ed\'erale de Lausanne (EPFL), CH-1015 Lausanne, Switzerland}

\author{Julian Schuhmacher}
\affiliation{\affiliationibm}
\affiliation{\affiliationepfl}
\author{Alberto Baiardi}
\affiliation{\affiliationibm}
\author{Francesco Tacchino}
\affiliation{\affiliationibm}
\author{Ivano Tavernelli}
\affiliation{\affiliationibm}

\begin{abstract}
Hybrid tensor networks offer a promising route to enhance the expressivity of classical tensor network methods by incorporating quantum states prepared on a quantum computer. 
Existing approaches are limited by the variational optimization of the quantum component of the tensor network.
In this work, we introduce an alternative strategy that combines a non-parametric quantum state prepared through quantum annealing and a classical isometric tensor network.
The latter is variationally optimized while the former is used as a fixed, boundary tensor resource in the form of classical shadows.
We demonstrate the feasibility of this approach through extensive numerical simulations on the transverse-field Ising model, showing that the optimization procedure remains robust under statistical and hardware noise.
Moreover, our results indicate that our newly proposed setup improves the accuracy of the obtained ground state approximation compared to the original quantum simulation, without increasing the depth of the applied quantum circuits.
Therefore, this setup offers a practical route to scale variational quantum algorithms towards the quantum utility scale.
\end{abstract}

\maketitle

\section{Introduction}
Efficiently representing the relevant quantum wavefunctions of many-body systems remains a central challenge in condensed-matter and high-energy physics. 
Among the most powerful classical tools developed for this purpose are tensor networks (TNs)~\cite{white1992density,fannes1992finitely,ostlund1995thermodynamic,verstraete2004renormalization,shi2006classical,vidal2008class,evenbly2014class,vidal2004efficient,schollwock2011density,orus2014practical,orus2019tensor,cirac2021matrix}, which provide compact and physically motivated representations of complex quantum states. 
Depending on their topology, tensor networks can accurately capture the entanglement structure characteristic to the ground and low-energy states of specific Hamiltonians~\cite{evenbly2011tensor,eisert2010colloquium}.
For example, Matrix Product States (MPSs) can efficiently encode low-energy eigenstates of one-dimensional gapped systems~\cite{verstraete2006matrix,hastings2007area,eisert2010colloquium}.
However, the computational cost of optimizing tensor networks gets prohibitively expensive when studying two- or higher-dimensional quantum systems~\cite{schuch2007computational,haferkamp2020contracting}.
This limits the expressivity of the ans\"atze that can be simulated in practice.
A promising approach to mitigate such limitations is the integration of quantum states into TNs, an framework known as hybrid tensor networks~\cite{yuan2021quantum,schuhmacher2025hybrid}.

Hybrid tensor networks leverage the idea that quantum states prepared on a quantum computer can, in principle, represent quantum wave functions too complex to be represented classically.
Therefore, such a hybrid setup clearly enhances the network expressivity compared to its purely classical counterpart.
However, practical applications of hybrid tensor networks face two major limitations that prevent scaling them to large system sizes~\cite{schuhmacher2025hybrid}.
First, when applied to variational problems, quantum tensors are often represented through parametrized quantum circuits.
Unfortunately, their optimization suffers from well-known trainability issues, restricting their applicability to relatively small numbers of qubits~\cite{mcclean2018barren,wang2021noise,thanasilp2023subtleties}.
Second, the most practical optimization strategies for hybrid tensor networks rely on a local, iterative procedure that requires, at each step, performing a partial quantum state tomography of the quantum tensors~\cite{schuhmacher2025hybrid}. 
This results in a considerable overhead in quantum resources, scaling linearly with the number of quantum registers associated with each tensor.

In this work, we introduce a novel approach that addresses both aspects.
Specifically, we propose to replace the variational quantum tensor with a non-parametric state prepared via quantum annealing~\cite{finnila1994quantum,kadowaki1998quantum,farhi2001quantum,albash2018adiabatic}, while confining the parametrized components to the classical parts of the TN.
A similar idea has been explored for time-evolution circuits combined with neural networks in~\cite{gentinetta2025correcting}.
Since such a setup does not involve any parametrized quantum circuit, the variational optimization does not require any iterative optimization of quantum circuits.
Additionally, quantum annealing is based on Hamiltonian simulation, which is among the most well-studied subroutines in quantum computing, as well as one of the most suitable for large scale experiments on current quantum processors~\cite{kim2023evidence,yu2023simulating,farrell2024quantum,miessen2024benchmarking,chowdhury2024enhancing,cochran2025visualizing,haghshenas2025digital,farrell2025digital,zemlevskiy2025scalable,schuhmacher2025observation,chai2025resource}.
To ensure proper normalization of the ansatz, we employ an isometric tensor network for the classical part, the Multi-scale Entanglement Renormalization Ansatz (MERA)~\cite{vidal2008class,evenbly2009algorithms}.
In practice, the annealing quantum states serve as a resource of boundary tensors for the MERA. 
To extract the relevant quantum information without resorting to full state tomography, we employ classical shadows~\cite{huang2020predicting,acharya2021shadow}.
These have been recently demonstrated at scale on multiple quantum information processing platforms~\cite{fischer2024dynamical,vermersch2024many,fischer2025large}, and can easily be integrated into TN workflows via their tensor representation~\cite{GarciaPerez2022_Vilma,filippov2022matrix,Akhtar2023_ShadowTomography-TensorNetworks,filippov2023scalable,mangini2025low,votto2025learning}.

We validate our hybrid approach by performing ground state optimization of the transverse-field Ising model Hamiltonian for multiple system sizes.
Our extensive numerical simulations indicate that our setup consistently improves the accuracy of the final ground state approximation compared to quantum annealing alone, effectively matching the outcomes of slower and longer annealing schedules without increasing the quantum circuit depth.
Additionally, we show that the optimization of the classical tensor networks is robust against both statistical and hardware noise.
These findings suggest that our approach is a promising candidate for scaling variational quantum algorithms to even larger system sizes.

\section{Methods}
\label{sec:methods}

\subsection{Non-parametric quantum state preparation}
We prepare an approximation to the ground state of a target Hamiltonian using a digital quantum annealing (QA)~\cite{albash2018adiabatic} protocol.
The system is first initialized in the ground state of an Hamiltonian, which can be easily prepared on a quantum processor.
The Hamiltonian is then adiabatically transformed into the target one.
Assuming that the system remains in the instantaneous ground state throughout the evolution, this procedure prepares the ground-state of the final, target Hamiltonian.
Unlike approaches based on parametrized quantum circuits, which suffer from theoretical (e.g., barren plateaus~\cite{Larocca2025_BarrenReview}) and practical limitations to trainability, and can therefore only be successfully run for limited system sizes, QA in principle provides a direct, scalable alternative for ground-state preparation.

Here, we will study the transverse-field Ising Hamiltonian for which the corresponding time-dependent Hamiltonian is
\begin{equation}\label{eq:annealing_hamiltonian}
    \hat{H}(t) = J(t) \sum_{i} Z_i Z_{i+1} + \lambda(t) \sum_i X_i \,,
\end{equation}
where $J(t)$ and $\lambda(t)$ are time-dependent parameters.
We apply an annealing schedule with $J(t) = - t / t_{\text{final}}$ and $\lambda(t) = 1.0$ such that at the final time $t = t_\text{final}$ the system is at the critical point, $|J|=|\lambda|$.
For the initial state, we prepare the ground state of the second term in~Eq.\eqref{eq:annealing_hamiltonian} at $t = 0$, which corresponds to the product state $|-\rangle^{\otimes N}$, where $N$ is the number of sites.
The quantum circuit implementing the time-evolution operator for the time-dependent Hamiltonian given in~Eq.\eqref{eq:annealing_hamiltonian} via a second order Trotterization is provided in Appendix~\ref{app:annealing_circuit}.

\begin{figure}
    \centering
    \includegraphics[width=0.95\linewidth]{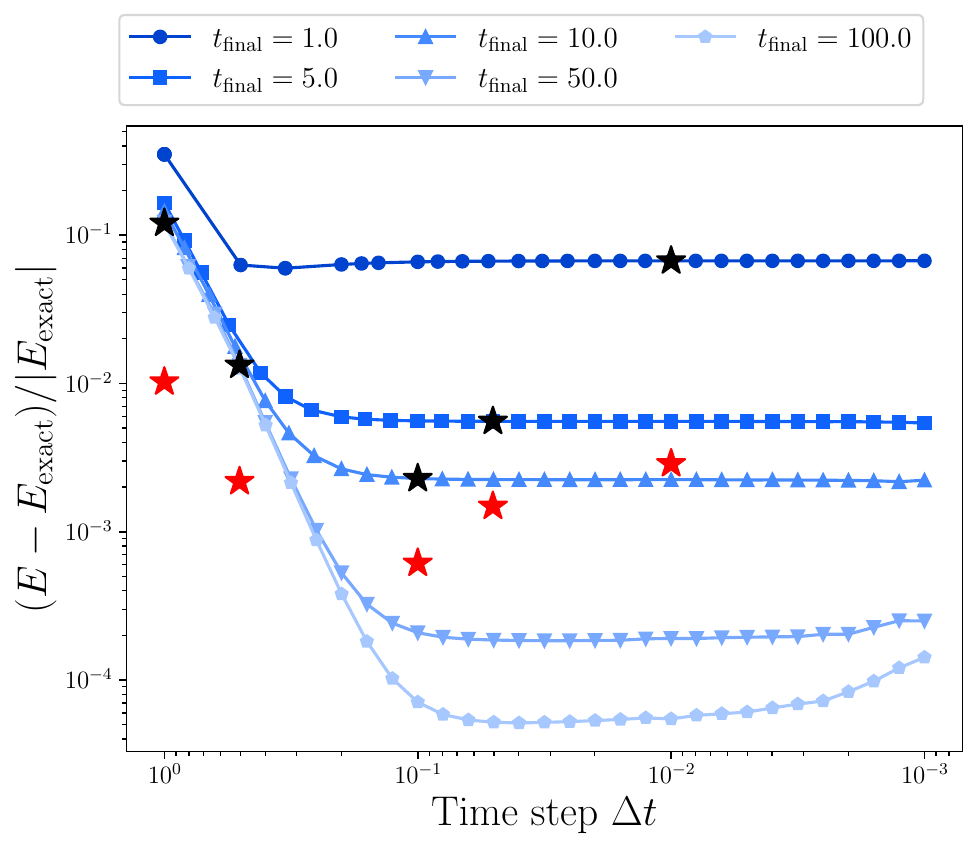}
    \caption{
        Relative energy error of a quantum circuit implementing the annealing for a system with $N=24$ sites via a Trotter decomposition.
        For large time steps $\Delta t$, the energy error is mainly due to the Trotter error.
        For small time steps $\Delta t$, the Trotter error is small enough such that the energy error saturates at a finite, limiting value due to the finite annealing time $t_\text{final}$.
        The black stars highlight the combination of final time $t_\text{final}$ and time-step $\Delta t$, that results in the quantum circuit with the closest two-qubit gate depth to 200.
        The red stars represent the error obtained using the quantum state obtained by using the resulting quantum circuit within our proposed hybrid MERA, and by subsequently optimizing it in a noiseless setting.
        The exact energy $E_\text{exact}$ is evaluated with exact diagonalization.
    }
    \label{fig:trotter_error}
\end{figure}

The accuracy of implementing the quantum annealing via a quantum circuit is affected by two error sources.
The first arises from discretizing the continuous evolution into a finite number of time steps through a Trotter factorization.
The second stems from the Trotter product formula decomposition~\cite{Miessen2023_QuantumDynamicsReview} applied within each individual time step.
In practice, these errors are amplified by the use of a finite $t_\text{final}$, since the ground state can, in principle, be reached only in the limit $t_\text{final} \rightarrow \infty$.
In figure~\ref{fig:trotter_error}, we compare the overall error for a system of $N=24$ sites for different annealing times $t_\text{final}$ and time steps $\Delta t$.
As expected, the Trotter error dominates for large $\Delta t$ values, and decreases with $\Delta t$.
For small $\Delta t$ values, the Trotter error becomes negligible, and the simulation accuracy is limited only by the finite annealing time $t_\text{final}$.
For each curve, the black star highlights, for a given final time $t_\text{final}$, the time step $\Delta t$ yielding the quantum circuit with a two-qubit gate depth closest to the reference value of 200 -- a reasonable estimate of the largest quantum circuits affordable for current generation of quantum processors.
This enables a direct comparison of the accuracy of the state prepared through QA for a given, fixed circuit depth budget and different combinations of $\Delta t$ and $t_\text{final}$.

\subsection{Hybrid MERA Tensor network}
The QA routine prepares, at the end of the annealing schedule, a quantum state that loosely approximates the ground state of the Hamiltonian.
To improve the quality of this state, we contract it with a Multiscale Entanglement Renormalization Ansatz (MERA) tensor network~\cite{yuan2021quantum}, \textit{i.e.} $\vert \Psi_\text{hMERA} \rangle = U_\mathrm{MERA} \vert \Psi_\text{QA} \rangle$.
We evaluate the energy $\langle \Psi_\text{hMERA} \vert \hat{H} \vert \Psi_\text{hMERA} \rangle$ by Heisenberg-propagating the Hamiltonian $\hat{H}$ with the classical tensor network, which therefore acts as a transformation of the underlying Hamiltonian
\begin{equation}\label{eq:transformed_hamiltonian}
    \hat{H}_\mathrm{MERA}(U_\mathrm{MERA}) = U_\mathrm{MERA}^\dagger \hat{H} U_\mathrm{MERA} \,.
\end{equation}
Notably, $U_{\mathrm{MERA}}$ may denote any operation that admits a classical tensor-network representation.
In our implementation, however, we specifically use the MERA~\cite{vidal2008class}, an isometric tensor network that preserves the Hamiltonian's eigenvalue spectrum. 
A second advantage of the MERA is that, when used to transform a local Hamiltonian $\hat{H}$ as in Eq.~\ref{eq:transformed_hamiltonian}, the locality of the unitarily-transformed operator $\hat{H}_\mathrm{MERA}$ does not grow uncontrollably.

\begin{figure}
    \centering
    \includegraphics[width=\linewidth]{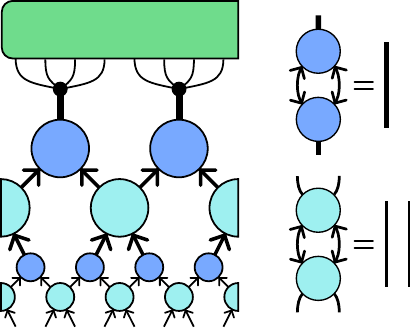}
    \caption{
        Hybrid MERA tensor network consisting of a quantum tensor (green), and two layers of the MERA.
        The black dots represent the fusion of the involved legs of the quantum tensor.
        The isometries in the MERA layers are highlighted in blue, while the unitaries are highlighted in teal.
        We use periodic MERA layers, which is depicted by the half-spheres for the unitaries at the edges.
        The contraction of the isometries and the unitaries with their respective complex conjugate over the legs with incoming arrows yields the identity, as shown in the figures on the right.
        The thickness of the bonds increases with their dimension.
    }
    \label{fig:hybrid_mera}
\end{figure}

The hybrid MERA structure employed in this work is shown in Figure~\ref{fig:hybrid_mera}.
The quantum state at the top is represented as a quantum tensor, and the classical TN is constructed from MERA layers.
Notably, we do not apply any truncations to the MERA, and therefore its bond dimension is $\chi_l = 2^{2^l}$ at the top of each layer $l$. 
Therefore, $n_l = \log_2 \chi_l = 2^l$ qubits are required to represent the indices of the $l$-th MERA layer, to be contracted with the quantum tensor.
For the example in Figure~\ref{fig:hybrid_mera} consisting of two classical MERA layers, four legs of the quantum tensor, each corresponding to a qubit, are fused together, represented by the black dot.
Given a fixed realization of the quantum tensor, the optimization of the MERA can be formally written as
\begin{equation}\label{eq:optimization}
    E_\mathrm{GS} = \min_{T \in \mathrm{MERA}} \langle \psi_\mathrm{QA}| \hat{H}_\mathrm{MERA}(T) |\psi_\mathrm{QA}\rangle \,,
\end{equation}
where the optimization is performed over all tensors $T$ in the MERA, $|\psi_\mathrm{QA} \rangle$ is the quantum state prepared with QA and $\hat{H}_\mathrm{MERA}$ is the transformed Hamiltonian in Eq.~\ref{eq:transformed_hamiltonian}).
To perform this optimization, we require a strategy to evaluate the expectation value of $\hat{H}_\mathrm{MERA}$ on $|\psi_\mathrm{QA}\rangle$.
The structure of $\hat{H}_\mathrm{MERA}$ (for instance, its decomposition in terms of Pauli strings) cannot be determined a priori and evolves instead as the optimization of the MERA proceeds.
Hence, the strategy must support generic operators.
In this scenario, quantum state tomography would allow for the evaluation of the relevant expectation values without repeated access to the quantum processor.
However, this procedure is exponentially costly, and the resulting quantum state representations cannot be effectively stored nor manipulated in classical form beyond small numbers of qubits.
We will address this issue below.
For the time being, let us assume that a valid representation of the quantum state is available in a form that can be contracted with the MERA layers, allowing us to calculate the quantities required for the optimization.

\subsubsection{Riemannian optimization of isometric tensor networks}
All the tensors entering the MERA layers are either isometries or unitaries. 
Therefore, they can be optimized with a global, gradient-based Riemannian optimization constrained on the manifold of isometric matrices $\mathcal{M}$~\cite{hauru2021riemannian,luchnikov2021riemannian,le2025riemannian}.
Together with the calculation of gradients via automatic differentiation~\cite{baydin2018automatic}, these techniques are the current state-of-the-art for the optimization of isometric tensor networks.

The following paragraph summarizes how these techniques can be used to update a single tensor $X \in \mathcal{M}$.
The extension to multiple tensors is trivial.
The loss function used for the optimization is the energy expectation value $E(X)$, calculated by contracting the Hamiltonian between the MERA and its complex conjugate (see Eq.~(\ref{eq:transformed_hamiltonian}) and then contracting the resulting tensor network with the prepared quantum state.
We denote the gradient of the energy with respect to variations of the tensor $X$ calculated with automatic differentiation as $\nabla E(X)$.
The Riemannian gradient is obtained by projecting this gradient to the tangent space of the manifold $\mathcal{M}$ at $X$, $\nabla_R E(x) \in T_X\mathcal{M}$.
The projection of a vector $Y \in \mathbb{C}^{n \times p}$ onto the tangent space $T_X\mathcal{M}$ can be calculated as~\cite{hauru2021riemannian}
\begin{equation}\label{eq:projection}
    Y' = Y - \frac{1}{2} X \left(X^\dagger Y + Y^\dagger X \right) \,,
\end{equation}
where $Y' \in T_X\mathcal{M}$.
We then use the Riemannian version of the ADAM optimizer~\cite{kingma2014adam} to compute the update of $X$~\cite{luchnikov2021riemannian,le2025riemannian}.
The main difference compared to standard ADAM is that, when calculating the momentum term, only elements belonging to the same tangent space $T_X\mathcal{M}$ can be summed together.
To this end, all momentum vectors must be brought into the correct tangent space via a parallel transport.
In practice, this corresponds to applying the projection defined in Eq.~\eqref{eq:projection} to the momentum vectors.
As a final step, the updated tensor $X' \in T_X\mathcal{M}$, which in general is not an isometry, is projected back to the manifold $\mathcal{M}$. 
We implement this projection step through a Singular Value Decomposition (SVD),
\begin{equation}
    X' = U S V^\dagger \rightarrow X'_\mathcal{M} = U V^\dagger \,,
\end{equation}
where $X'_\mathcal{M} \in \mathcal{M}$.
This process can straightforwardly be generalized to a product manifold of isometric tensors $\mathcal{M}^{\times n_\text{tensors}}$, where $n_\text{tensors}$ is the number of tensors in the applied TN.

\subsection{Energy estimation with classical shadows}\label{sec:ic_povms}

As mentioned above, a reconstruction of the annealing state via full tomography protocols would represent, in principle, the ideal way to transfer quantum resources into the classical MERA. 
However, such an approach is unfeasible for large quantum systems. 
Moreover, a tensor network representation of the resulting wavefunction would be computationally expensive, if not for those states that can be already efficiently described classically and do not require the use of a quantum processor.

To circumvent these limitations, we resort to the framework of classical shadows, a type of informationally-complete POVM~\cite{huang2020predicting,acharya2021shadow}. 
A brief review of the key concepts and equations is presented in Appendix~\ref{app:cs}. 
Crucially, for local Pauli shadows both the measurement operators and the shadows themselves can be straightforwardly represented as Matrix Product Operators with a bond dimension $b=1$.
Therefore, the quantum tensor at the top of the hybrid MERA in Figure~\ref{fig:hybrid_mera} can be represented as a sum over tensor-product MPOs, where the sum runs over the different shots collected from the quantum processor.
When calculating expectation values, each MPO is individually contracted with the MERA and the Hamiltonian before summing over all terms.

Because our energy estimates rely on local Pauli shadows, it is important to note that applying the MERA layers to the original Hamiltonian, as in Eq~\eqref{eq:transformed_hamiltonian}, generally alters its locality.
Here, locality refers to the Pauli weight of the elements in the Pauli decomposition of $\hat{H}$ and $\hat{H}_\mathrm{MERA}$. 
In the worst case, the locality increases as $k = O(2^l)$ where $l$ is the number of MERA layers.
Following Ref.~\cite{huang2020predicting}, the bound for the number of measurements $S$ required to estimate expectation values of a set of observables $\{O_i\}$ up to an additive error $\epsilon$ is then given by
\begin{equation}\label{eq:resources_mera}
    S \sim \mathcal{O}\left( \frac{\log{M}}{\epsilon^2} 16^l \max_{1\leq i \leq M} \lVert O_i \rVert^2_{\infty} \right) \,,
\end{equation}
where $\lVert \,\cdot\, \rVert_\infty$ denotes the operator infinity-norm.
The argumentation for this bound can is provided in appendix~\ref{app:locality}. 
For a random MERA, simulations clearly indicate that the variance increases with the number of layers.
Empirical evidence indicates that, for an optimized MERA, the variance of the energy estimator exhibits only a small increase. Consequently, the bound presented in Eq.~\eqref{eq:resources_mera} should be regarded as a loose bound. 
(see Appendix~\ref{app:pauli_weight}).

\section{Results}\label{sec:results}
First, we investigate the expressivity of our newly proposed \textit{ansatz}, \textit{i.e.} how effectively the quantum state obtained from the quantum annealing can be refined using the MERA.
At this stage, we consider an exact contraction of the annealing quantum state (represented, in our simulations, as an MPS) and the MERA. 
We will discuss the impact of shot noise at a later stage.
We use 1000 optimization steps in the Riemannian ADAM optimization of the MERA, with standard choices of $\alpha=0.01$, $\beta_1=0.9$, $\beta_2=0.999$ and $\epsilon=10^{-8}$ as hyperparameters.
For the $N=24$ site system, the red stars in Fig.~\ref{fig:trotter_error} indicate the energy obtained after the optimization of the hybrid MERA.
For a given $\Delta t$, these values should be compared to the corresponding energy of the state prepared through quantum annealing, \textit{i.e.} the initial point of the optimization, represented as black stars.
Encouragingly, we observe a consistent reduction in the energy error across all initial points.

\begin{figure}
    \centering
    \includegraphics[width=0.95\linewidth]{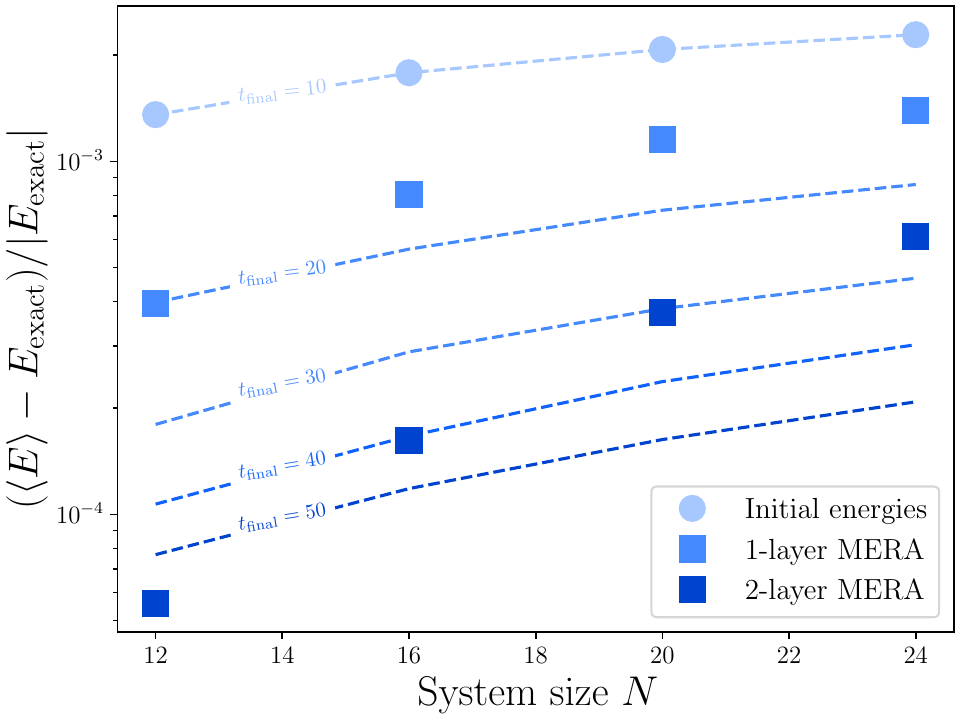}
    \caption{
        Ground-state energy obtained from quantum annealing (light blue circles) and subsequent improvement for a  MERA with 1 (blue squares) and 2 (dark blue squares) layers for different system sizes $N$.
        The dashed line represent the energy obtained from quantum annealing with different final times $t_\text{final}$.
    }
    \label{fig:system_size}
\end{figure}

For all subsequent simulations, we select $t_\mathrm{final} = 10$ and $\Delta t = 0.1$, as this set of parameters is associated with the QA schedule yielding the lowest-energy state.
When this circuit is used to prepare the initial state across different system sizes, the subsequent MERA optimization consistently reduces the energy error, as shown in Fig.~\ref{fig:system_size}. 
In the figure, the dots denote the initial QA energies, while the squares indicate the final hybrid MERA values. 
The dashed lines show the energies obtained by increasing the final time of the annealing to $t_\mathrm{final} = \{20, 30, 40, 50\}$.
Across all systems sizes, the MERA effectively reduces the required total annealing time -- and, therefore, the depth of the corresponding quantum circuit -- to achieve a certain energy accuracy.
For $N = 12$, the improvement factor is larger than 5 and remains above 2 even for $N=24$.

\begin{figure}
    \centering
    \includegraphics[width=0.95\linewidth]{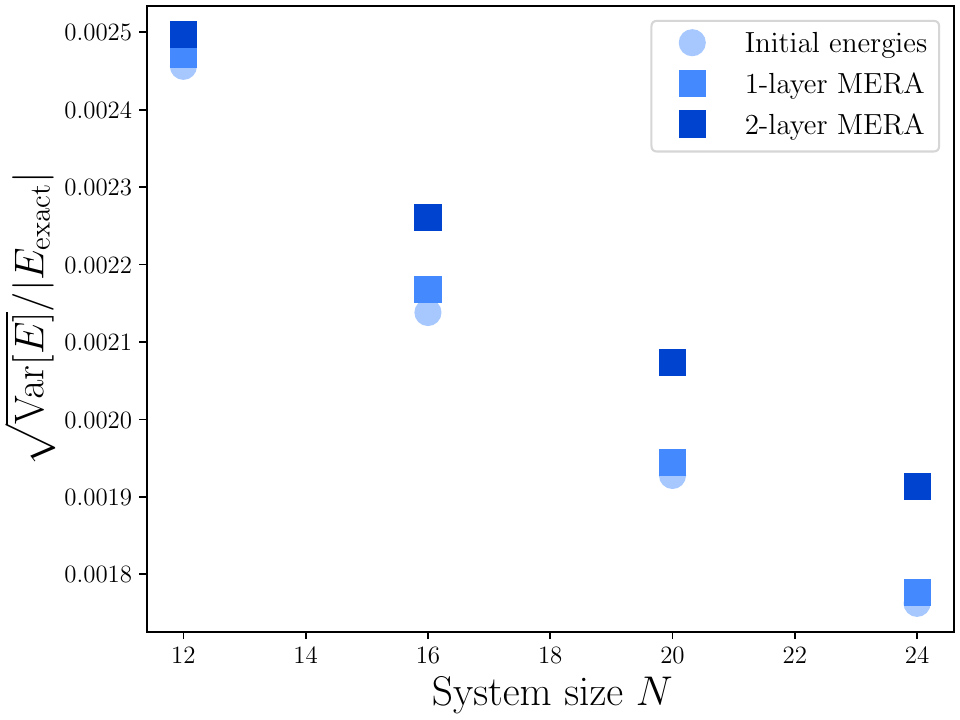}
    \caption{Variance of the energies obtained in figure~\ref{fig:system_size} evaluated with $S=10^5$ POVM samples for the state generated with QA (light blue circle), and for the same state combined with an optimized MERA with 1 (blue squares) and 2 (dark blue squares) layers.}
    \label{fig:variance}
\end{figure}

\medskip
We now extend the analysis reported in Fig.~\ref{fig:system_size} by accounting for the impact of shot noise in the accuracy of the quantum-classical interface.
To this end, we first simulate a classical shadows data acquisition step with $S = 10^5$ samples and evaluate the variance of the resulting energy estimator. 
The result is shown in Fig.~\ref{fig:variance}.
Interestingly, although revolving the MERA into the Hamiltonian as in Eq.~(\ref{eq:transformed_hamiltonian}) in principle increases its non-locality, this does not manifest in practice when the transformation is calculated based on the \textit{optimized} MERA.
In fact, the variance does not increase significantly in our experiments compared to the original Hamiltonian.
In Appendix~\ref{app:pauli_weight}, we study this behaviour in more detail.

\begin{figure}
    \centering
    \includegraphics[width=1.0\linewidth]{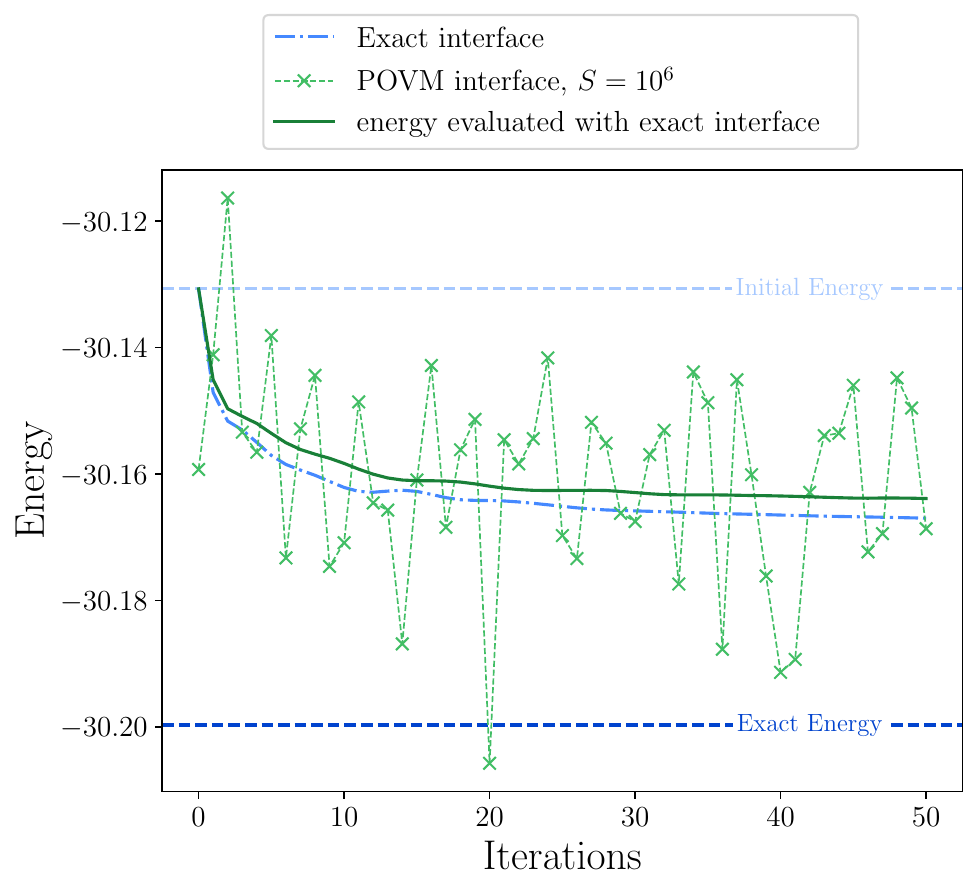}
    \caption{
        Optimization of a hybrid MERA applying a classical shadow interface to extract a classical representation of the QA state, and to contract it with the classical MERA layers.
        The dashed light (dark) blue line shows the QA (exact) energy.
        The dash-dotted blue line shows the energy optimization of the hybrid MERA when the interface is implemented exactly.
        The green lines show the energy optimization of the hybrid MERA when the classical shadow interface is used.
        For reference, the green solid lines show there energy of the hybrid MERA evaluated with the exact interface, using the MERA parameters obtained with the classical shadows.
    }
    \label{fig:povm_optimization}
\end{figure}

Based on the evaluated variance, we can now estimate the number of shots $S$ required to resolve a certain energy difference, assuming that the variance scales as $O(1 / S)$ (i.e., the statistical error follows an inverse square-root law), as discussed in Appendix~\ref{app:povm_resource_estimation}.
For the 24-site system, this extrapolation indicates that $S \simeq 10^6$ snapshots are needed to resolve the difference between the initial energy guess and the exact ground-state energy of the system.
To verify whether this number of shots yields a stable optimization trajectory, we simulate the full shot-based optimization of the 24-site system, shown in Fig.~\ref{fig:povm_optimization}.
To prevent that the resulting energy estimator is biased towards the used set of POVM snapshots, we use an independent set of $S$ shadows for its evaluation~\cite{filippov2022matrix}. 
Additionally, we resample $S$ shadows at each optimization step to obtain an unbiased estimator for the Riemannian gradients.
However, notice that the whole dataset can, in principle, be collected at the beginning of the experiment (i.e., without accessing the quantum processor after the data acquisition phase), since the annealing quantum state does not depend on the MERA itself.

The energy evolution of the hybrid MERA optimization using classical shadows is shown as the green crosses.
For reference, the green solid line shows the energy of the hybrid MERA evaluated with the exact quantum state wavefunction but using the MERA parameters obtained under shot noise.
The blue dotted line shows the corresponding optimization using the exact contraction without statistical effects.
We also recall here that a longer, fully-converged optimization with an exact interface and 1000 iterations was reported in Fig.~\ref{fig:system_size}.

The green solid line converges smoothly, indicating that the MERA parameters continuously and consistently yield lower energies even though the corresponding energy estimate oscillates.
These results suggest that accurate MERA parameters can be obtained by optimizing them using a relatively limited number of classical snapshots of the annealing state, even if that yields a relatively high energy variance.
A higher number of snapshots can then be used only at the final step to obtain the last energy estimate with a higher precision.

\medskip

\begin{figure*}
    \centering
    \includegraphics[width=1.0\linewidth]{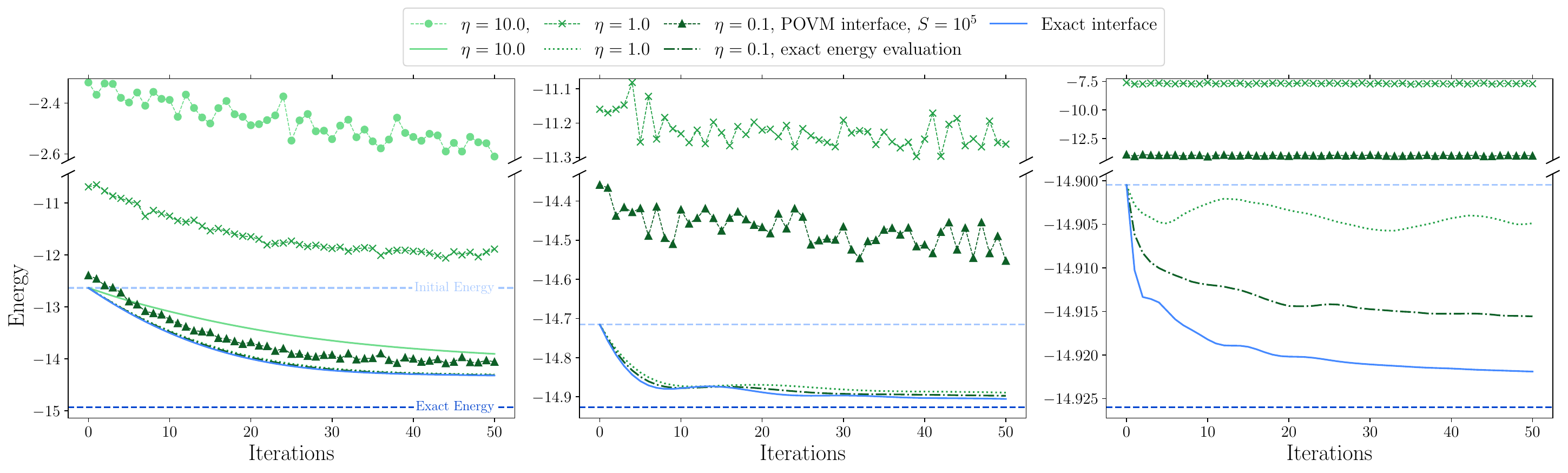}
    \caption{
        Optimization of a hybrid MERA using a classical shadows interface while accounting for hardware noise.
        The initial state is prepared with an annealing circuit consisting of 10 (left), 20 (middle), and 50 (right) Trotter steps with time steps $\Delta t = 1.0, 0.5$ and $0.2$, respectively.
        The green curves are obtained using the noise profile of \texttt{ibm\_marrakesh} scaled with the corresponding factor $\eta$.
        The circles, crosses and triangles represent the energies obtained with the classical shadows interface.
        The solid, dotted and dash-dotted lines correspond to the energies evaluated with the exact interface, using the MERA parameters obtained with the classical shadows.
        The solid blue line corresponds to a reference optimization where both the MERA parameters and the energies are obtained via the exact interface.
        The horizontal dashed lines show the initial (light-blue) and exact (dark-blue) energies, evaluated using the noiseless QA state and exact diagonalization, respectively.
    }
    \label{fig:hardware_noise}
\end{figure*}

Finally, we investigate how hardware noise affects the optimization of our hybrid MERA ansatz. 
To this end, we simulate the optimization for a 12-site system based on the noise parameters obtained from a noise characterization of the IBM Quantum \texttt{ibm\_marrakesh} device.
We prepare the initial states using an annealing schedule with final time $t_\text{final} = 10$ consisting of 10, 20 and 50 Trotter steps.
We sample $S = 10^5$ classical shadows at each optimization step and analyse the impact of varying noise strengths $\eta = 10.0, 1.0$ and $0.1$ on the simulation accuracy.
The noise strength $\eta$ represents the ratio of the error rates in the simulated noise model compared to the original one (see Appendix~\ref{app:noise_model} for details).
The results are shown in Fig.~\ref{fig:hardware_noise}.

The green circles, crosses and triangles represent the hybrid MERA optimization where the MERA parameters and the energies are obtained via a noisy shadows interface (i.e., including statistical and hardware noise) for different noise strengths $\eta=10.0, 1.0, 0.1$, respectively.
The solid, dotted and dash-dotted green lines represent the hybrid MERA optimization where the MERA parameters are obtained using the noisy shadows interface, but the energies are then evaluated using the exact wavefunction of the QA state contracted with the MERA.
The solid blue line corresponds to a reference optimization where both the MERA parameters and the energies are obtained via the exact interface.
Finally, the horizontal dashed lines show the initial (light-blue) and exact (dark-blue) energies, evaluated with the noiseless annealing state and exact diagonalization, respectively.

For the state generated using 10 Trotter steps (figure on the left), the hybrid MERA optimization consistently improves upon the energy of the noisy annealing state for all noise strengths $\eta$.
While for $\eta = 10.0$ (circles) and $\eta = 1.0$ (crosses) the energy remains higher than the energy of the noiseless annealing state (horizontal light-blue, dashed line), for $\eta = 0.1$ (triangles) the energy converges to a lower value.
Surprisingly, the hybrid MERA optimization trajectories where the MERA parameters are obtained with classical shadows and the energies are evaluated using exact wavefunctions (green lines) yield energies comparable to the hybrid MERA optimization where also the MERA parameters are obtained via the exact interface (solid blue line).
For $\eta = 1.0$ (dotted green line) and $\eta = 0.1$ (dash-dotted green line), the two results closely match, whereas for the $\eta = 10.0$ case (solid green line), the convergence of the energy is noticeably slower.

For the two other initial states (20 and 50 Trotter steps in the annealing circuit), we only simulate $\eta = 1.0$ and $\eta = 0.1$.
For 20 Trotter steps (figure in the middle), the results are similar to the case with 10 Trotter steps.
Both curves improve the energy estimate with respect to the noisy annealing state (crosses and triangles), and the contraction of the optimized MERA with the exact annealing state (dotted and dash-dotted green lines) results in energies close to the exact optimization (solid blue line).
The behaviours begin to diverge for the optimization with the 50 Trotter step initial state (figure on the right).
There, the improvement in the energy of the hybrid MERA is minimal (crosses and triangles), and the contraction of the optimized MERA with the exact initial state (dotted and dash-dotted green lines) starts to deviate more and more from the exact optimization (solid blue line) as the noise strength increases.

The obtained results suggest that, even in the presence of statistical and hardware noise, the optimization of the MERA parameters converge to values yielding accurate energy estimates.
Hence, the MERA layers can be safely optimized with a relatively cheap (in terms of sampling and error handling costs) noisy approximation of the quantum state.
A high quality ground state energy estimate can then be retrieved at the last step using the resulting MERA parameters and a better approximation of the quantum state (e.g., using a high number of snapshots and applying error mitigation).

\section{Conclusions}\label{sec:conclusion}
We have introduced a hybrid tensor network architecture that combines non-parametric quantum state preparation via quantum annealing with a classical MERA tensor network. 
This setup leverages classical shadow measurements to circumvent the limitations of earlier proposals~\cite{schuhmacher2025hybrid}, and allows for reliable ground state approximations of quantum many-body systems without resorting to parametrized quantum circuits.

Our numerical experiments on the transverse-field Ising model demonstrate that our new hybrid ansatz improves the accuracy of the target ground state approximation, effectively emulating a longer annealing time, and therefore deeper circuits.
This represents an interesting, equilibrium physics counterpart of recent proposals for effective operator backpropagation in quantum dynamics and generic quantum circuit simulation~\cite{Rudolph2025_PP-Review,Miller2025_MajoranaPropagation,Nys2025_MajoranaPropagation,Fuller2025_OBP}.
Additionally, we observed through numerical simulations that the classical optimization of the tensor network remains robust under statistical and hardware noise. 
In many cases, the parameters in the MERA converge to values that are comparable to those obtained in exact, noiseless benchmarks.
This approach yields an optimization routine for hybrid MERA wavefunctions which requires minimal quantum resources per optimization iteration, and allows us to execute a highly-accurate quantum calculation only at the end of the optimization.

The accuracy of our newly proposed hybrid scheme is inherently constrained by the quality of the initial quantum state, which is not parametrized  (apart from the total annealing time and the discretization parameters) by design. 
While this choice bypasses potential trainability issues, it also limits the expressiveness of the state and, consequently, the quality of the final result. 
Incorporating a limited number of variational parameters in the circuit used for the quantum state preparation could enhance the flexibility of our method and further improve its accuracy.
Moreover, a more effective use of quantum resources can further enhance the performance of the algorithm.
For instance, some of the classical shadows collected from the quantum processor could be reused across multiple optimization steps to reduce the overall measurement overhead. 
However, this strategy requires a careful calibration, as reusing snapshots will inevitably lead to biased energy and gradient estimators.
Moreover, while classical shadows are basis-independent and thus well-suited for an iterative optimization of a Hamiltonian that is updated at every iteration, tailoring the POVM measurement scheme to the structure of the transformed Hamiltonian could reduce the variance of the energy estimator~\cite{garcia2021learning,fischer2024dual,Caprotti2024_DualOptimization,mangini2025low,korhonen2025improving}.
Noteworthy, regarding the classical tensor network component, the approach is not limited to MERA and can be easily generalized to other isometric tensor networks. 
The gradient evaluation imposes practical constraints on the bond dimensions, requiring a careful balance between expressiveness and computational feasibility.
However, provided that the state prepared on the quantum computer is sufficiently close to the exact, target one, a compact classical TN remains achievable.

\section*{Acknowledgements}
We thank Laurin Fischer for helpful discussions regarding the POVMs. This research was supported by the project RESQUE (Rethinking Quantum Simulations in the Quantum Utility Era, grant number 20QU-1\_225229) and by the National Center of Competence in
Research (NCCR) SPIN (grant number 225153), both funded by the Swiss National Science Foundation. IBM, the IBM logo, and ibm.com are trademarks of International Business Machines Corp., registered in many jurisdictions worldwide. Other product and service names might be trademarks of IBM or other companies. The current list of IBM trademarks is available at \url{https://www.ibm.com/legal/copytrade}.

\appendix

\section{Trotter circuit for quantum annealing}
\label{app:annealing_circuit}

\begin{figure}
    \centering
    \begin{subfigure}[c]{0.25\linewidth}
        \caption{}
        \includegraphics[width=\linewidth]{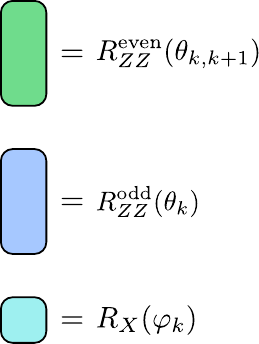}
    \end{subfigure}
    \hfill
    \begin{subfigure}[c]{0.70\linewidth}
        \caption{}
        \includegraphics[width=\linewidth]{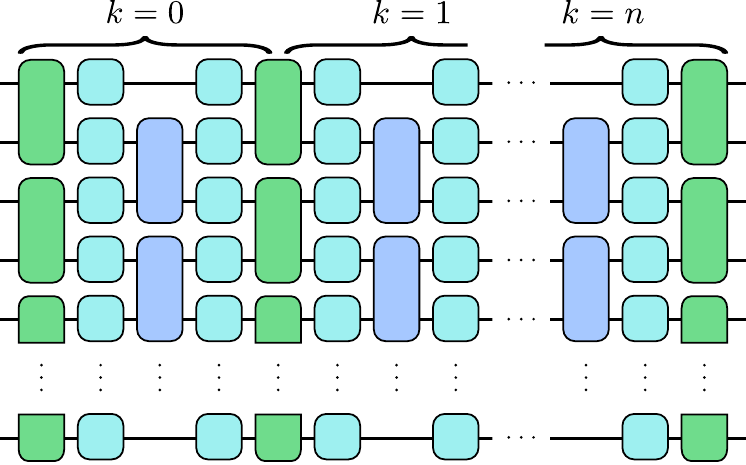}
    \end{subfigure}
    \caption{
        Quantum circuit implementing the quantum annealing via a second order Trotter evolution of the transverse-field Ising Hamiltonian in equation~\eqref{eq:annealing_hamiltonian}.
        \textbf{a} shows the three building blocks of the circuit. 
        The $R_{ZZ}$ gates on the even bonds contain the angles $\theta^e_{k,k+1}$ defined in the main text. 
        The even $R_{ZZ}$ gates in the first and last layer are only half rotations with the angle $\theta^e_k$ defined in the main text.
        The $R_{ZZ}$ gates on the odd bonds have the angles $\theta^o_k = 2 \, J(t_k + \frac{\Delta t}{2}) \Delta t$.
        The $R_X$ gates are rotations with the angle $\varphi_k = 2 \, \lambda(t_k + \frac{\Delta t}{2}) \frac{\Delta t}{2}$.
        \textbf{b} Shows the overall structure of the quantum circuit.
    }
    \label{fig:annealing_circuit}
\end{figure}

The time-evolution operator associated with the Hamiltonian in equation~\eqref{eq:annealing_hamiltonian} is given by the time-ordered exponential~\cite{hatano2005finding}
\begin{equation}\label{eq:time_ordered_exponential}
    G(t_2; t_1) = T\left[ \exp \left( -i \int_{t_1}^{t_2} \hat{H}(s) \, ds \right) \right] \,,
\end{equation}
where $T$ denotes the time ordering operator.
The Hamiltonian can be decomposed into three non-commuting terms
\begin{align}\label{eq:non_commuting_terms}
    A(t) &= J(t) \sum_{i \, \text{odd}} Z_i Z_{i+1} \,, \notag \\
    B(t) &= \lambda(t) \sum_i X_i \,, \\
    C(t) &= J(t) \sum_{i \, \text{even}} Z_i Z_{i+1} \notag \,,
\end{align}
such that a propagator $G(t + \Delta t; t)$ for a time step $\Delta t$ can be implemented with the second-order Trotter decomposition as
\begin{align}\label{eq:second_order_trotter}
    G_2(t + \Delta t; t) = \, & e^{-i \frac{\Delta t}{2} C(t + \frac{\Delta t}{2})} e^{-i \frac{\Delta t}{2} B(t + \frac{\Delta t}{2})} \notag \\
                              & e^{-i \Delta t A(t + \frac{\Delta t}{2})} e^{-i \frac{\Delta t}{2} B(t + \frac{\Delta t}{2})} \\ 
                              & e^{-i \frac{\Delta t}{2} C(t + \frac{\Delta t}{2})} \,. \notag
\end{align}
The two-qubit gate term $C(t)$, which appears at the beginning and at the end of a given time step, can be merged across consecutive time steps.
This reduces the number of two-qubit gate layers in the quantum circuit that implements the time-evolution.
Specifically, the evolution of the term $C(t)$ is implemented with a layer of $R_{ZZ}(\theta)$ gates with the angle $\theta^e_k = 2 \, J(t_k + \frac{\Delta t}{2}) \frac{\Delta t}{2}$, where $t_k = k \Delta t$ is the time after $k$ time steps. 
Since the time dependence is only contained in the parameters, the evolution of $C(t)$ of consecutive time-steps can be combined in a single layer of $R_{ZZ}$ gates with the angle $\theta^e_{k,k+1} = 2 \left[J(t_k + \frac{\Delta t}{2}) + J(t_{k+1} + \frac{\Delta t}{2})\right] \frac{\Delta t}{2}$.
The quantum circuit implementing the time evolution is shown in figure~\ref{fig:annealing_circuit}.

\section{Classical shadows}\label{app:cs}

A POVM consists of a set of $m$ positive semidefinite Hermitian operators $M = \{ M_k \}_{k \in \{1, ..., m\}}$ which satisfy the relation $\sum_{k=1}^{m} M_k = \mathbb{I}$~\cite{fischer2024dual}.
If this set of operators is \textit{informationally complete} (i.e. it spans the space of Hermitian operators), there exists a set of dual operators $D = \{D_k\}_{k \in \{1, ..., m\}}$, such that any observable $O$ can be decomposed as
\begin{equation}
    O = \sum_{k=1}^{m} \operatorname{Tr}[O D_k] M_k = \sum_{k=1}^{m} \omega_k M_k \,,
\end{equation}
where $\omega_k = \operatorname{Tr}[O D_k]$ are the operator weights.
Given this decomposition of $O$ and a quantum state $\rho$, the expectation value $\langle O\rangle_\rho$ is obtained as
\begin{equation}
    \langle O\rangle_\rho = \operatorname{Tr}[\rho O] = \sum_{k} \omega_k \,\operatorname{Tr}[\rho M_k] = \sum_k \omega_k p_k \,,
\end{equation}
where $p_k = \operatorname{Tr}[\rho M_k]$ is the probability of observing outcome $k$.
Given $S$ measurement outcomes $\{k^{(1)}, \ldots, k^{(S)}\}$, we can construct an unbiased Monte-Carlo estimator $\langle \hat{o} \rangle$ of $\langle O \rangle_{\rho}$ as
\begin{equation}
    \langle \hat{o} \rangle = \frac{1}{S} \sum_{s=1}^{S} \omega_{k^{(s)}}.
\end{equation}
The standard error on the mean of this estimator is
\begin{align}\label{eq:variance}
    \operatorname{Var}[\hat{o}] &= \frac{1}{S}\frac{1}{S-1} \sum_{s=1}^{S} \left(\omega_{k^{(s)}} - \langle \hat{o} \rangle \right)^2 \,.
\end{align}

\section{Locality of MERA-transformed Hamiltonians}
\label{app:locality}

\begin{figure}
    \centering
    \includegraphics[width=1.0\linewidth]{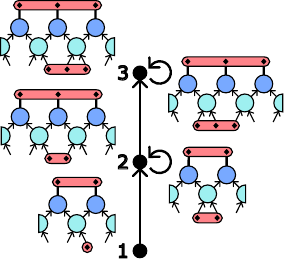}
    \caption{
        Transitions between the locality of operators, when starting from one- or two-local operators.
        The possible transitions are given by the black arrows.
        The diagrams show how these transitions are realized.
        The locality of the operators is given in terms of the number of links it spans at the bottom or the top of the MERA layer.
        The maximal locality that can be reached through these transitions is three links.
        In terms of qubits that are required to represent these links, the worst-case locality is $3\cdot 2^l$, since $2^l$ qubits are required to implement a single link at the top of layer $l$ in the MERA.
    }
    \label{fig:locality}
\end{figure}

To predict $M$ expectation values $\operatorname{Tr}[O_i \rho]$ up to additive error $\epsilon$ for arbitrary $k$-local observables $O_i$, the required number of measurements is~\cite{huang2020predicting}
\begin{equation}\label{eq:resources_classical_shadows}
    S \sim \mathcal{O}\left( \frac{\log{M}}{\epsilon^2} 4^k \max_{1\leq i \leq M} \lVert O_i \rVert^2_{\infty} \right) \,,
\end{equation}
where $\lVert \,\cdot\, \rVert_\infty$ denotes the operator infinity-norm.
Transforming the Hamiltonian $\hat{H}$ with MERA layers as in equation~\eqref{eq:transformed_hamiltonian} changes the locality of the terms in the Hamiltonian.
When starting from one- and two-local operators, which are the main building blocks of many lattice Hamiltonians, there are a limited number of ways how the locality of these operator can change when Heisenberg-propagated through a layer of the MERA.
All possibilities are shown in figure~\ref{fig:locality}, where the transitions in the locality can be $1\rightarrow2$, $2\rightarrow2$, $2\rightarrow3$ or $3\rightarrow3$, where the first number denotes the number of links spanned by the operator before it is propagated through the layer (operator at the bottom), and the second number is the number of links spanned after the operator is propagated.
Therefore, the transformed operators can span at most three links when propagated through a layer of the MERA.
The number of qubits representing a link at the top of layer $l$ is $2^l$.
Therefore, in the worst case, the locality of the operators expressed in terms of qubits also increases as $k = O(2^l)$, and the resource estimate in equation~\eqref{eq:resources_classical_shadows} becomes
\begin{equation}
    S \sim \mathcal{O}\left( \frac{\log{M}}{\epsilon^2} 16^l \max_{1\leq i \leq M} \lVert O_i \rVert^2_{\infty} \right) \,.
\end{equation}

\section{Variance of MERA-transformed Hamiltonians}\label{app:pauli_weight}

In the following, we investigate how much Pauli strings with different weight contribute to the variance of the energy estimator of the Hamiltonian transformed through backpropagation of the MERA tensor network.
For this, we split the transformed Hamiltonian $\hat{H}$ into groups of Pauli strings with the same weight $w$,
\begin{equation}\label{eq:operator}
    \hat{H} = \sum_w \sum_{\substack{i \\ |P_i| = w}} c_i P_i = \sum_w \hat{P}_w \,,
\end{equation}
where $|\cdot|$ denotes the weight of the Pauli string and $\hat{P}_w$ is the operator obtained by summing together all Pauli strings of weight $w$.
The variance of $\hat{H}$ can be written as~\cite{fischer2024dual}
\begin{equation}\label{eq:povm_variance}
    \operatorname{Var}(\hat{H}) = \frac{1}{N-1}\left[ \sum_k p_k \omega_k^2 - \left(\sum_k p_k \omega_k\right)^2 \right] \,,
\end{equation}
where $S$, $p_k$ and $\omega_k$ are defined in section~\ref{app:cs}.
Substituting equation~\eqref{eq:operator} into equation~\eqref{eq:povm_variance} leads to (note that we do not explicitly include the prefactor in favour of clarity)

\begin{widetext}
\begin{align}
    \operatorname{Var}(\hat{H}) &\propto \sum_k p_k \left(\sum_w \operatorname{Tr}[D_k \hat{P}_w]\right)^2 - \left( \sum_k p_k \sum_w \operatorname{Tr}[D_k \hat{P}_w] \right)^2 \notag\\
                                &= \sum_w \left[\sum_k p_k \left(\omega_k^{(w)}\right)^2 - \left(\sum_k p_k \omega_k^{(w)} \right)^2\right] + 2 \sum_{w' > w} \left[ \sum_k p_k \omega_k^{(w)} \omega_k^{(w')} - \left(\sum_k p_k \omega_k^{(w)}\right) \left( \sum_{k'} p_{k'} \omega_{k'}^{(w')} \right) \right] \notag\\
                                &= \sum_w \operatorname{Var}(\hat{P}_w) + 2 \sum_{w' > w} \operatorname{Cov}(\hat{P}_w, \hat{P}_{w'}) \,,
\end{align}
\end{widetext}
where in the first line we have use the definition of $\omega_k$ and $\hat{H}$, in the second line we have defined $\omega_k^{(w)} = \operatorname{Tr}[D_k \hat{P}_w]$ and expanded and rearranged the expressions, and in the last line we have used the definitions of the variance and the covariance.
We can now define the contribution of the Pauli operators $\hat{P}_w$ of a given weight $w$ to the variance as
\begin{equation}
    C_w = \operatorname{Var}(\hat{P}_w) + \sum_{w' \neq w} \operatorname{Cov}(\hat{P}_w,\hat{P}_{w'}) \,,
\end{equation}
such that $\operatorname{Var}(\hat{H}) = \sum_w C_w$.

\begin{figure}
    \centering
    \includegraphics[width=1.0\linewidth]{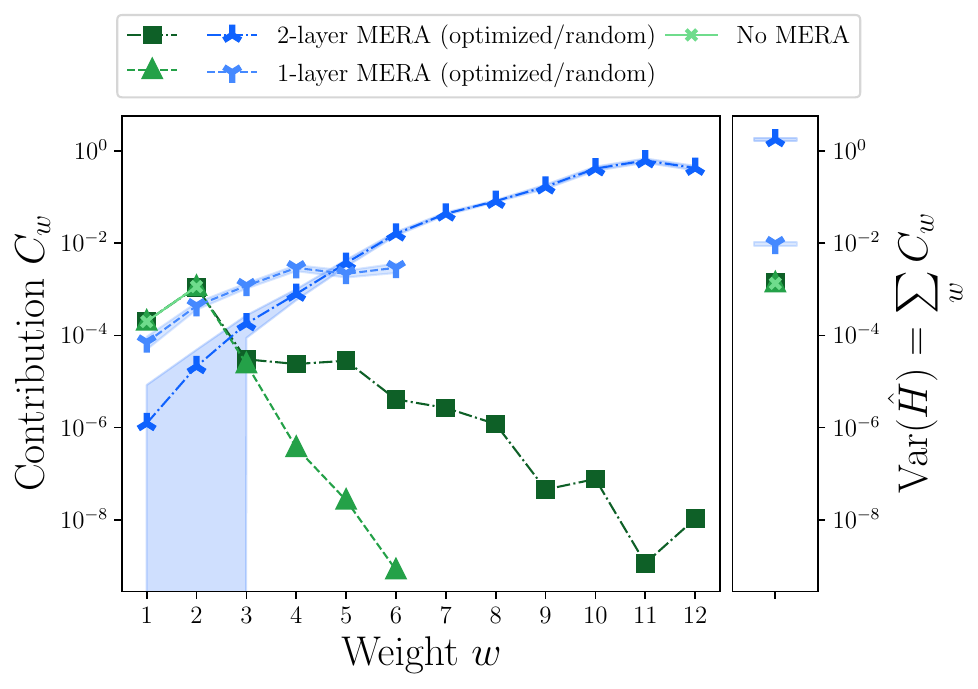}
    \caption{Contribution to variance resolved by weight of Pauli strings comprising the observable for different options for the MERA layers.}
    \label{fig:variance_by_pauli_weight}
\end{figure}

In figure~\ref{fig:variance_by_pauli_weight}, we report the contribution $C_w$ to the variance for a 12-site TFIM where the Hamiltonian is transformed by different choices of the MERA. 
The individual contributions $C_w$ are shown in the panel on the left for the original Hamiltonian (green crosses), and the Hamiltonian transformed with an optimized one-layer MERA (green triangles), an optimized two-layer MERA (green squares), a random one-layer MERA (blue downward facing triangular crosses), and a random two-layer MERA (blue upward facing triangular crosses).
The panel on the right shows the corresponding total variance $\mathrm{Var}(\hat H)$.
For the random MERA, the data points are obtained as the average over 100 instances of the variance estimation.
The highlighted region represents their standard deviation.
All quantities are calculated based on sets of $10^5$ POVM snapshots. 

In the case of the random MERA, the variance is dominated by the contributions associated with high-weight Pauli operators.
Conversely, for the optimized MERA, the high-weight Pauli operators only have a minor contribution.
Together with the observation that the MERA parameters can be reliably optimized even in a noisy setting (see section~\ref{sec:results}), this suggests that the overhead associated with measuring the transformed Hamiltonian does not increase significantly compared to the original Hamiltonian, although the former is in principle less local than the latter.

\section{Scaling of the noise model}
\label{app:noise_model}

The noise models applied in the noisy simulations include readout, depolarization and relaxation errors.
To construct a noise model for which the effective noise is scaled by a factor of $\eta$, we adjust the error sources as follows:
\begin{itemize}
    \item Readout errors $e_\text{ro}$:
    \begin{equation*}
        \tilde{e}_{ro} = \eta \,e_\text{ro}
    \end{equation*}
    \item Depolarization error rates $p_\text{gate}$:
    \begin{equation*}
        \tilde{p}_\text{gate} = \eta \, p_\text{gate}
    \end{equation*}
    \item Relaxation errors: These errors are governed by T1 (decay towards the $|0\rangle$ state) and T2 (decay towards the maximally mixed state). To obtain an effective noise which is scaled by $\eta$, we scale the T1 and T2 times by the factor
    \begin{equation*}
        \alpha = - \frac{t}{t_g} \log\left[1 - \eta \left( 1 - e^{-\frac{t_g}{t}} \right) \right] \,
    \end{equation*}
    where $t$ is either the T1 or T2 time, and $t_g$ is the longest duration of the implemented gates (excluding measurements and resets).
\end{itemize}

For the noisy simulations, we use the factors $\eta = 10.0, 1.0$ and $0.1$.
While the factors $\eta = 10.0$ and $\eta = 1.0$ correspond to realistic noise model, it is not necessarily true for the one obtained with $\eta = 0.1$. 
For the readout error rates and the gate errors, a factor of $\eta = 0.1$ results in a hardware with a noise profile that is targeted for error correction. 
However, for the T1 and T2 times, the scaling with $\eta = 0.1$ results in an increase of the times by a factor of $\alpha \approx 100$.
Such a large increase is unrealistic.
In practice, however, the same effect as increasing T1/T2 times can be obtained by simultaneously increasing T1/T2 times and reducing the gate durations.
Therefore, we believe that the simulated noise model with $\eta = 0.1$ is a good representation of next the generation of quantum hardware~\cite{mohseni2024build}.

\section{Resource estimation for POVM interface}
\label{app:povm_resource_estimation}
For the POVM-based optimization, we can estimate the number of POVM snapshots required to obtain accurate energy estimates as follows.
To resolve an energy scale $\Delta E$, we require an estimator with variance $\operatorname{Var}_S$, evaluated using $S$ snapshots, such that
\begin{equation}\label{eq:energy_scale}
    \sqrt{\operatorname{Var}_S} \approx f\, \Delta E \,,
\end{equation}
where $f \in \mathbb{R}$ controls the desired statistical resolution.

In the main text, we evaluated the variance of the energy estimator using $S' = 10^5$ POVM snapshots. 
Assuming the usual $O(1/S)$ scaling of the variance, the corresponding single-shot variance can be approximated as
\begin{equation}
    \operatorname{Var}_1 \approx \operatorname{Var}_{S'}\, S' \,.
\end{equation}
Using the same scaling, the variance for an arbitrary number of snapshots $S$ is given by
\begin{equation}\label{eq:variance_S}
    \operatorname{Var}_S = \frac{\operatorname{Var}_1}{S}
    = \frac{\operatorname{Var}_{S'}\, S'}{S} \,.
\end{equation}
This further assumes that the variance does not change significantly over the course of the optimization, which is supported by the numerical evidence presented in the main text.
Combining Eq.~\eqref{eq:energy_scale} with Eq.~\eqref{eq:variance_S}, we obtain
\begin{equation}
    S \approx \frac{\operatorname{Var}_{S'}\, S'}{(f\, \Delta E)^2} \,,
\end{equation}
as an estimate for the number of snapshots required to resolve an energy scale $f\,\Delta E$.

For the POVM-based optimization of the $N=24$ site TFIM with $l = 2$ MERA layers, the energy difference between the initial and the exact energy is $\Delta E \approx 0.07$ (see Fig.~\ref{fig:system_size}), and the estimated variance is $\operatorname{Var}_{10^5} \approx 0.0033$ (see Fig.~\ref{fig:variance}).
Choosing $f = 1/4$, such that statistical fluctuations typically remain within $\Delta E$, we obtain $S \approx 1.06 \cdot 10^6$.
Accordingly, we use $S = 10^6$ POVM snapshots in the POVM-based optimization shown in Fig.~\ref{fig:povm_optimization}.

\section{POVM-induced bias in MERA optimization}

When classical shadows are used both to estimate energies and to compute gradients, correlations induced by the reuse of POVM data can lead to violations of the variational principle. 
In the following, we examine the different sources of bias in the resulting energy estimators due to finite statistics and overfitting.

\begin{figure*}[ht]
    \centering
    \begin{subfigure}[b]{0.48\textwidth}
        \centering
        \caption{}
        \includegraphics[width=\linewidth]{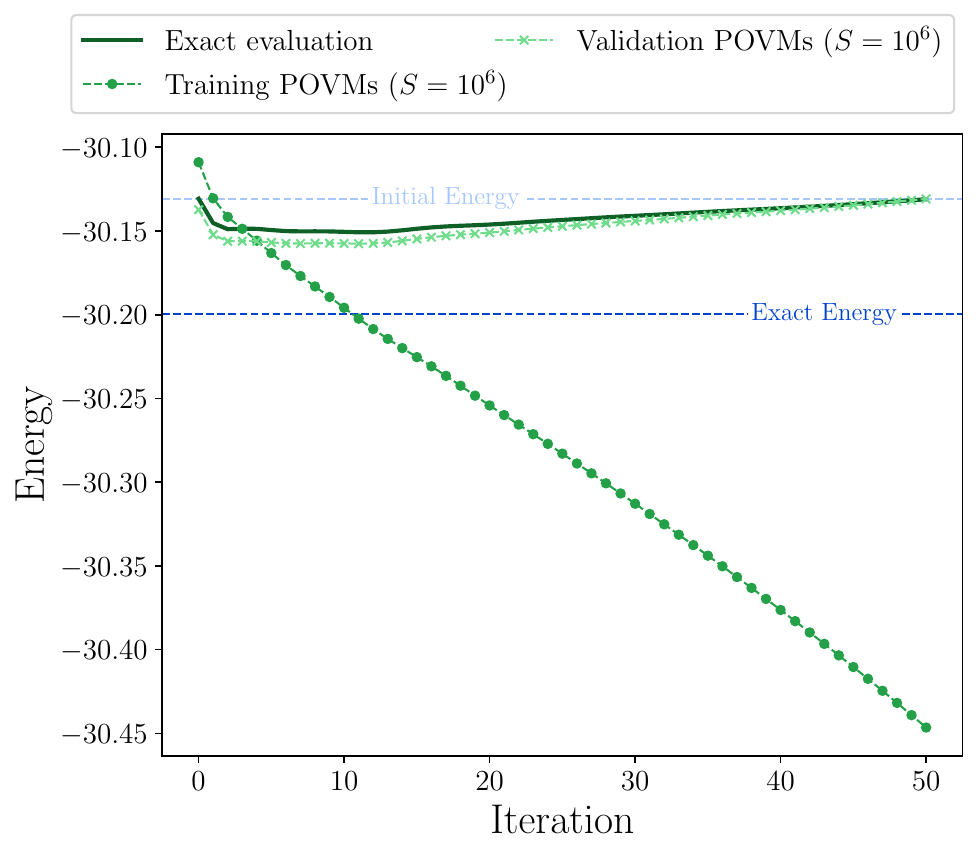}
        \label{fig:biased_optimization}
    \end{subfigure}%
    \hfill
    \begin{subfigure}[b]{0.48\textwidth}
        \centering
        \caption{}
        \includegraphics[width=\linewidth]{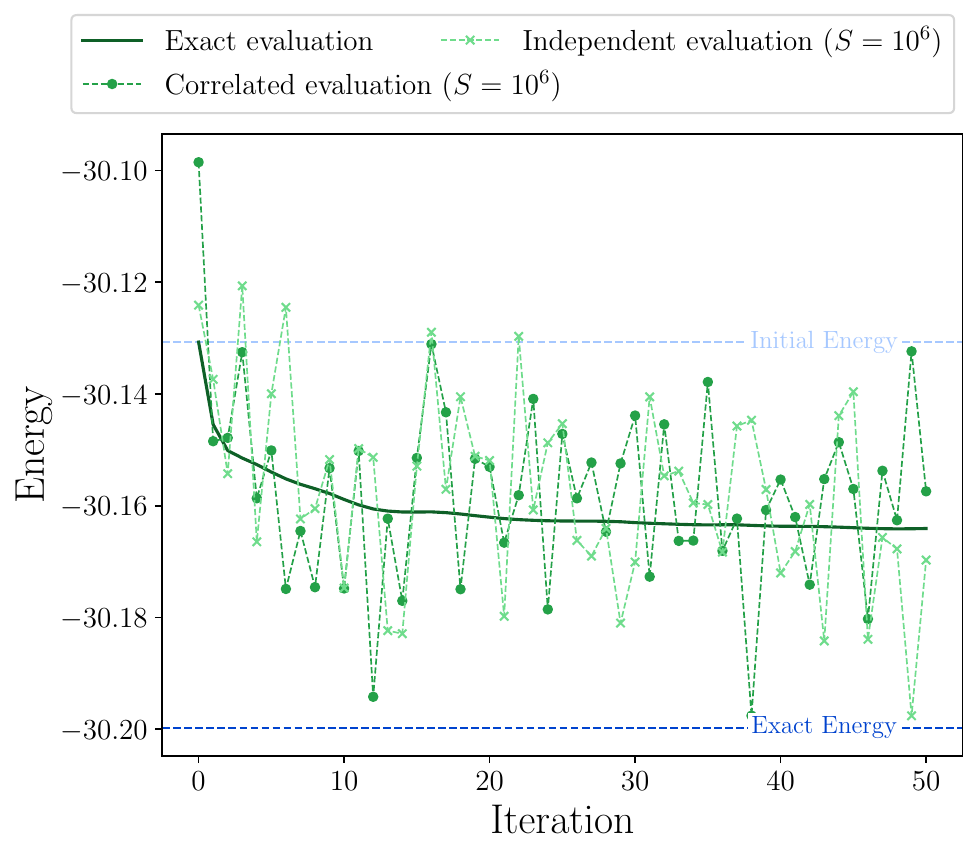}
        \label{fig:unbiased_optimization}
    \end{subfigure}
    \caption{
        Bias study for different POVM sampling protocols.
        (a) Optimization using a fixed set of POVM snapshots for gradient estimation (green dots) leads to a systematic decrease of the empirical energy below the exact ground-state energy (blue dashed line), indicating overfitting to the specific snapshot realization. 
        Evaluating the same parameters on an independent set of POVMs (light green crosses) removes this evaluation bias.
        (b) Comparison of correlated and independent evaluations when POVM snapshots are resampled at each optimization step. 
        Using the same snapshots for gradient and energy evaluation (green dots) introduces instantaneous correlations that lead to slightly lower energy estimates, while an independent evaluation (light green crosses) yields unbiased energies. Since snapshots are refreshed at each step, these correlations do not accumulate, and the residual bias remains within statistical uncertainty.
    }
    \label{fig:bias_study}
\end{figure*}

We consider several optimization protocols that differ in how POVM snapshots are used to evaluate gradients and energies.
These protocols can be viewed as successively removing different sources of bias.
Specifically, we study the following cases and report their comparison in Fig.~\ref{fig:bias_study}:
\begin{itemize}
    \item[\textit{(i)}] A single fixed set of POVM snapshots is used throughout the optimization to estimate the Riemannian gradients and to evaluate the energy (green dots in Fig.~\ref{fig:biased_optimization}).
    \item[\textit{(ii)}] Gradients are computed using a fixed set of POVM snapshots, while the energy is evaluated on a different fixed set (light green crosses in Fig.~\ref{fig:biased_optimization}).
    \item[\textit{(iii)}] A new set of POVM snapshots is resampled at each optimization step and used both to estimate the gradients and to evaluate the energy. The energy is evaluated after updating the MERA parameters (green dots in Fig.~\ref{fig:unbiased_optimization}).
    \item[\textit{(iv)}] Gradients and energies are each evaluated on independent, newly resampled sets of POVMs at every optimization step (light green crosses in Fig~\ref{fig:unbiased_optimization}).
\end{itemize}

In case \textit{(i)}, the optimization effectively minimizes a surrogate objective defined by the specific realization of the POVM data. 
As a result, the MERA parameters converge to values that minimize the empirical energy estimator rather than the true energy. 
This manifests itself in a decrease of the energy below the exact ground-state energy when evaluated on the same pool of snapshots.

Case \textit{(ii)} eliminates this evaluation bias in the energy by using an independent set of POVM snapshots for its evaluation, analogous to a validation dataset in a machine-learning context.
While this yields an unbiased evaluation of the energy, the parameters of the MERA remain biased due to overfitting to the training POVM data, as reflected in the increased energy when evaluated independently.

Using newly resampled POVMs for the gradient estimation at each optimization step prevents this form of overfitting, as illustrated by cases \textit{(iii)} and \textit{(iv)}.
Case \textit{(iii)} serves as an intermediate step that isolates the effect of correlations within a single optimization step, without allowing these correlations to accumulate over time.
In this setting, the use of identical POVM data for gradient and energy estimation induces a residual correlation between the energy estimator and the updated MERA parameters within each step, which can lead to a bias towards lower energies.
However, since the POVM samples are refreshed at every step, these correlations do not accumulate over time and are significantly reduced compared to case \textit{(i)}.
Empirically, this bias appears to be smaller than the statistical uncertainty arising from the POVM sampling process.

In case \textit{(iv)}, gradients and energies are evaluated on independent POVM samples, ensuring that the energy estimator is unbiased with respect to the current MERA parameters.
The resulting optimization is therefore variational up to statistical fluctuations.
This protocol is employed in the main text.

\bibliography{references}

\end{document}